\documentclass[12pt, draftclsnofoot, onecolumn]{IEEEtran}

\ifCLASSOPTIONcompsoc
  \usepackage[nocompress]{cite}
\else
  \usepackage{cite}
\fi

\usepackage[utf8]{inputenc}
\usepackage{mathrsfs}
\usepackage{tabularx,lipsum,environ,amsmath,amssymb}
\usepackage[boxruled,linesnumbered]{algorithm2e}
\usepackage{cleveref}
\crefname{Lemma}{Lemma}{Lemmas}
\usepackage{booktabs}
\usepackage{algorithm2e}
\usepackage[table,xcdraw]{xcolor}
\usepackage{tabularx,lipsum,environ,amsmath,amssymb}

\usepackage{algorithm2e}
\usepackage{graphicx}            
\usepackage{amsmath}             
\usepackage{amssymb}
\usepackage{array}
\usepackage{layout}
\usepackage[margin=1.04in]{geometry}
\usepackage{amsmath,amssymb}

\makeatletter
\newsavebox\myboxA
\newsavebox\myboxB
\newlength\mylenA

\newcommand*\xoverline[2][0.75]{%
	\sbox{\myboxA}{$\m@th#2$}%
	\setbox\myboxB\null
	\ht\myboxB=\ht\myboxA%
	\dp\myboxB=\dp\myboxA%
	\wd\myboxB=#1\wd\myboxA
	\sbox\myboxB{$\m@th\overline{\copy\myboxB}$}
	\setlength\mylenA{\the\wd\myboxA}
	\addtolength\mylenA{-\the\wd\myboxB}%
	\ifdim\wd\myboxB<\wd\myboxA%
	\rlap{\hskip 0.5\mylenA\usebox\myboxB}{\usebox\myboxA}%
	\else
	\hskip -0.5\mylenA\rlap{\usebox\myboxA}{\hskip 0.5\mylenA\usebox\myboxB}%
	\fi}
\makeatother

\usepackage{etoolbox}

\makeatletter
\patchcmd{\@maketitle}{\raggedright}{\centering}{}{}
\patchcmd{\@maketitle}{\raggedright}{\centering}{}{}
\makeatother

\begin{document}

\title{Memory-assisted Statistically-ranked RF Beam Training Algorithm for Sparse MIMO}

\author{\IEEEauthorblockN{Krishan K. Tiwari$^1$, \textit{SMIEEE}, Eckhard Grass$^{1,2}$, John S. Thompson$^4$, \textit{FIEEE},\\ and Rolf Kraemer$^{1,3}$ }\\
	\IEEEauthorblockA{$^1$IHP -- Leibniz-Institut f\"{u}r innovative Mikroelektronik, Im Technologiepark 25, 15236, Frankfurt (Oder), Germany.\\$^2$Humboldt-Universit\"{a}t zu Berlin, 10099, Berlin, Germany.~~~~~\\$^3$BTU-Cottbus, 03046, Cottbus, Germany.\\ $^4$IDCOM, School of Engineering, Edinburgh, EH9 3JL, U.K. \\
		Email addresses: {tiwari, grass, kraemer}@ihp-microelectronics.com, John.Thompson@ed.ac.uk}
}


\maketitle

\begin{abstract}

This paper presents a novel radio frequency (RF) beam training algorithm for sparse multiple input multiple output (MIMO) channels using unitary RF beamforming codebooks at transmitter (Tx) and receiver (Rx). The algorithm leverages statistical knowledge from past beam data for expedited beam search with statistically-minimal training overheads. Beams are tested in the order of their ranks based on their probabilities for providing a communication link. For low beam entropy scenarios, statistically-ranked beam search performs excellent in reducing the average number of beam tests per Tx-Rx beam pair identification for a communication link. For high beam entropy cases, a hybrid algorithm involving both memory-assisted statistically-ranked (MarS) beam search and multi-level (ML) beam search is also proposed. Savings in training overheads increase with decrease in beam entropy and increase in MIMO channel dimensions. 



\end{abstract}

\IEEEpeerreviewmaketitle

\section{Introduction}

For sparse MIMO channels, it is easier to learn the channel in beamspace than in the spatial signal space \cite{sayeed} --\cite{bradsong} as the number of available multi-path components (MPCs) for communications is much smaller compared to MIMO channel dimensions. In other words, the MIMO channel matrix $H$ is rank deficient with its rank being much smaller than its spatial dimensions. In such scenarios, RF beamforming allows low cost, low power consumption, and miniaturized hardware implementation as compared to fully digital baseband processing by avoiding extra RF chains \cite{mainpap}, \cite{mt}. The beamforming weights are implemented using digitally-controlled, constant-modulus analog phase-shifters at RF stage or intermediate-frequency (IF) stage. RF beamforming also corresponds to the first stage of the hybrid beamforming for sparse MIMO channels \cite{mainpap}, \cite{mt}.

Sparsity in beamspace reduces the problem of channel learning to identification of the best Tx-Rx beam combination in the received signal strength indication (RSSI) sense. RSSI is obtained by taking the magnitude of the receiver output. Because of limited Tx power, higher noise floor due to larger bandwidth, and propagation conditions, only a few beam combinations can establish a communication link for a given range.

Two prominent techniques for searching the best Tx-Rx beam combinations are: i) Exhaustive beam search and ii) Hierarchical beam search: this method is also known as `multi-level' (ML) beam search or `tree-based' beam search \cite{mlpaper2}. In exhaustive beam search, as the name suggests, all possible combinations of Tx and Rx beams are tested. The Tx-Rx beam pair yielding maximum RSSI is selected for payload data communication. This method has the disadvantage of very high training overheads since all possible Tx-Rx beam combinations need to be tested. For ML method, the beam search is performed in multiple levels. Progressively narrower beams are employed in multiple levels of beam training to search the best RSSI Tx-Rx beam combination only for the angular sector identified by the preceding level beam search \cite{mlpaper2}, \cite{ccwc}. The ML beam training algorithm cuts down the total number of beam tests needed for identification of the best Tx-Rx beam combination. This not only saves beam training overheads but also reduces mean communication loss due to false-beam selections on account of receiver noise \cite{ccwc}, \cite{wcsp}. However, ML beam search suffers from the disadvantage that the beam gains are lower for initial beam search levels as smaller number of antenna elements are switched on for beamforming for broader beams. This can be a major limitation for low signal to noise ratio (SNR) scenarios at larger ranges or with limited effective isotropic radiated powers (EIRPs).


The carbon footprint of information and communication technology (ICT) systems was as large as that of global air travel in 2008 \cite{energy}. The deployment of ICT systems has been growing since then and is likely to grow further in coming years. It is important to save beam training time and thereby power consumption as much as possible. Therefore, it is imperative to minimize the number of Tx-Rx beam tests and at the same time use high-gain narrow beams to realize communication links over larger distances. This calls for using beamforming with all antenna elements to provide narrow and high gain beams even during channel learning phases similar to exhaustive beam search. At the same time, it is also important to minimize the number of beam tests for minimum training overheads and maximum efficiency. 

In this paper, we present a novel RF beam training algorithm which leverages the statistical information, extracted from past beam training data, to achieve successful beam search with statistically-minimum overheads. We call the algorithm as memory-assisted statistically-ranked (MarS) beam search algorithm. The method works well for low beam entropy cases, i.e., when the beam probabilities are distributed very unevenly. The beam entropy $E$ is defined by equation (\ref{entropy}), 
\begin{equation}
\label{entropy}
\centering
E=-\sum_{{i=1}}^{N}p_ilog (p_i),
\end{equation}
where $p_i$ is the probability of the $i^{th}$ beam to provide a communication path and $N$ is the number of Tx or Rx beams. The lower the beam randomness, the lower is the beam entropy $E$. The best Tx-Rx beam combinations can be identified in a few initial tests with very high success rates if the Tx and Rx beam entropies are low due to low randomness of the channel. For high beam entropy cases, i.e., when the beam probability mass function (PMF) is closer to uniform distribution and the number of beams is large, we propose a hybrid beam search method involving both MarS and ML techniques. The initial level broader beams are obtained by combining later-level narrower beams such that the initial-level broader beams have as low entropy as possible. The proposed algorithms also help faster beam-based localization by quick Tx-Rx beam alignment. These techniques will also be useful for beam alignment for mm-wave / THz backhaul links in 5G or beyond 5G networks.

The rest of the paper is organized as follows: Section \ref{prob} specifies the problem statement, Section \ref{mars} describes the MarS algorithm, Section \ref{hybrid} presents the hybrid beam training algorithm, Section \ref{matlab} presents Matlab Monte Carlo simulation results, and Section \ref{conc} summarizes and concludes the paper. 

The following notations have been used: $\boldsymbol{x}$ is a vector, $x$ is a scalar, $\otimes$ denotes Kronecker product, length$(\boldsymbol{x})$ is the number of elements in vector $\boldsymbol{x}$, $\odot$ denotes Hadamard product, sum ({$\boldsymbol{x}$) denotes sum of all elements in vector $\boldsymbol{x}$, and $\boldsymbol{x}=[1:n]$ denotes a vector $\boldsymbol{x}$ given by first $n$ natural numbers from $1$ to $n$.

\section{Problem Statement}
\label{prob}
ML or exhaustive beam search algorithms are suitable for situations in which all beam directions are highly likely to be equiprobable, e.g., cellular applications due to very random locations of fast-moving users and/or scatterers. In other words, these algorithms do not take into account the probabilities for different beams to provide a communication link by corresponding to a MPC. 

At mm-wave and THz frequencies, the sparsity can cause the MIMO channel to be limited to the line-of-sight (LoS) component, i.e., the specular path, e.g., indoor radio channel for the virtual reality (VR) use case of EU H2020 WORTECS project [Section 2.2 of \cite{wvr}]. For the VR use case, we note that all the beam directions are not equally likely because of the following reasons: i) The user typically wears the VR devices and starts at designated locations in the VR room. ii) the users have limited mobility at a speed of 4 km/h, unlike extremely fast-moving users for cellular applications. iii) scatterers also have no or only limited mobility. iv) users' movement profiles are typically governed by a VR programme. v) Most of the times, movements to VR room extremities are not expected or will be for extremely short durations during the course of VR experiences. Therefore, different Tx and Rx beams will have different probabilities for establishing a communication link.
  
It is desirable to leverage the non-uniform PMF of Tx/Rx beams for the purpose of faster beam training, i.e., to  exploit the statistical knowledge of the MPCs to expedite the channel estimation process via identification of the best Tx-Rx beam combination as detailed in sections \ref{mars} and \ref{hybrid}. 

Without any loss of generality, we consider orthogonal or discrete Fourier transform (DFT) beamforming codebooks at Tx and Rx for minimal inter-beam coupling and with uniform linear array (ULA) implementation because the VR users move on a planar surface [Section 2 of \cite{bt}].

\section{Memory-assisted Statistically-ranked RF Beam Training Algorithm} 
\label{mars}
We consider one-dimensional (1D) beamforming in the azimuth plane, i.e., a ULA implementation for the VR use case and also for simplicity. 2D beamforming is an obvious and straightforward extension for planar array implementations. 

The whole beam training space can be divided into angular/beam grids using DFT beamforming codebooks at Tx and Rx. If the radio path aligns exactly with main response axes (MRAs) of Tx and Rx beams, then maximum beamforming gain $G=N_{TX} \times N_{RX}$ will be realised, where $N_{TX}$ and $N_{RX}$ are the numbers of Tx and Rx antenna elements, respectively. If there are angular offsets between the radio path and beam MRAs, then beam scalloping/cusping loss will be observed. Beam cusping decreases spectral efficiency by a statistical mean value of 1 bps/Hz [Section 3.9 of \cite{mt}].

Even though the proposed algorithms will be more effective with very large number of Tx and Rx antenna elements, we consider smaller MIMO channel dimensions for the ease of representation in this paper without any loss of generality. As illustrated in Figure \ref{bs} for a toy example, the beams can be labeled. Typically, a mobile user equipment (UE), e.g., a VR head mounted display (HMD), has relatively limited space. Without any loss of generality, in this paper we take Tx to be at VR access point (AP) with larger number of antenna elements and Rx to be at UE with smaller number of antenna elements.


\begin{figure}[!t]
	\hspace{-18pt}
	\includegraphics[width=7in, scale=1.9]{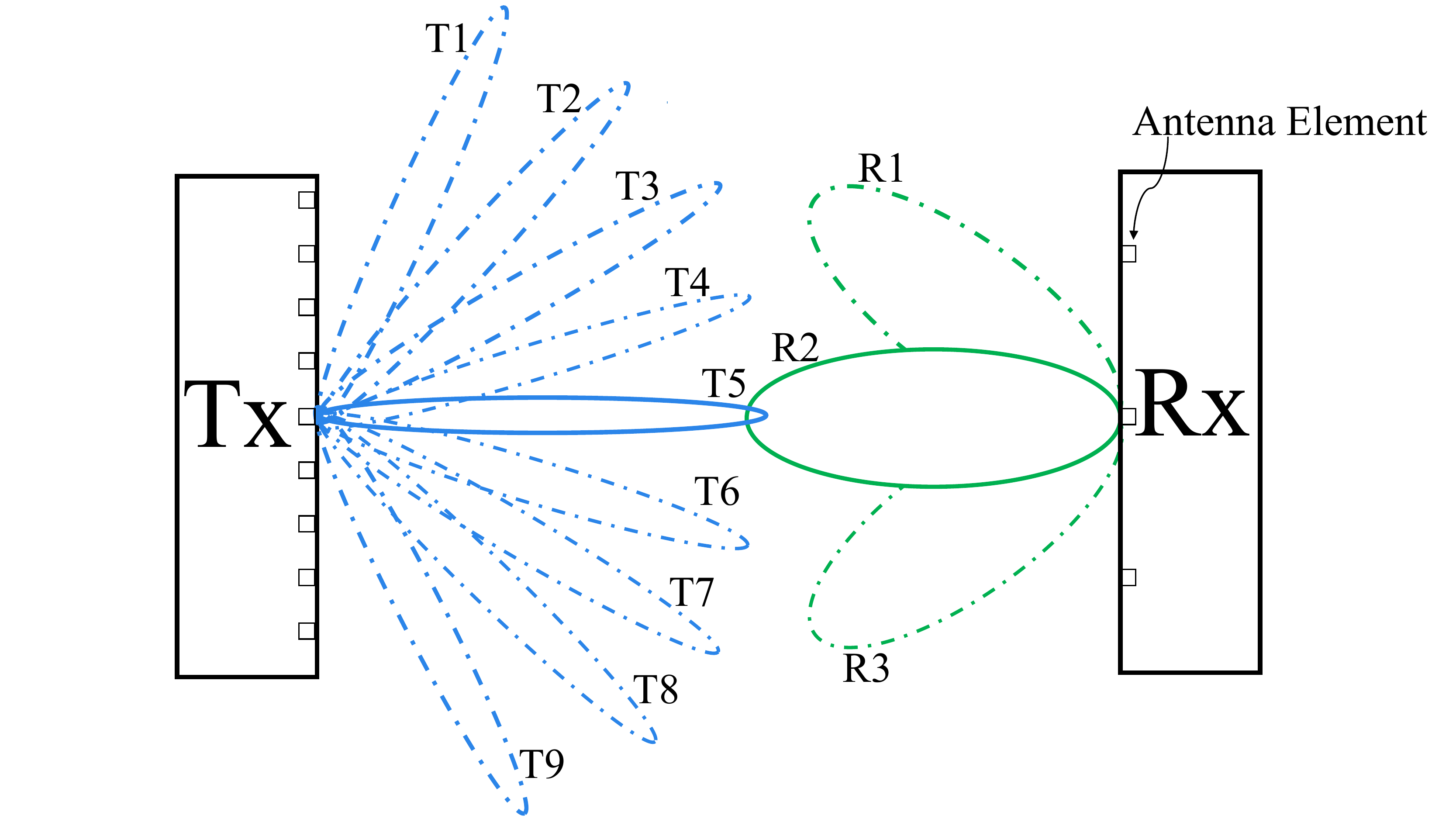}
	\caption{Labelled Tx and Rx beams}
	\label{bs}
\end{figure} 

In Figure \ref{bs}, the central Tx beams, i.e., T4, T5, and T6 and Rx beam R2 are more likely to correspond to the actual MPC (LoS for our case) for communication since the users are more likely to be in the central zone of the room than in the extremities on an average. We assume that the users are likely to begin from the centre of the room. The users will be free to rotate around, so we assume Rx beam statistics to be independent of Tx beam statistics.\footnote{In cases of correlation between Tx and Rx beams, the correlation can be used advantageously to further reduce training time and overheads.} Tx beam PMF will be more non-uniform than Rx beam PMF due to smaller number of Rx beams and free rotation of the user, thereby of the HMD. The beams can be ranked according to their statistics, with beams having higher success probability getting higher ranks. Arbitrary PMFs of Tx and Rx beams, along with corresponding rankings, have been considered in Tables \ref{Tx Beam Rankings} and \ref{Rx Beam Rankings}, respectively. 

\begin{table}[ht]
	\caption{Low entropy Tx Beam PMF and rankings} 
	\centering 
	\begin{tabular}{c c c c} 
		\hline\hline 
		Sl.\# & Beam\# & Probability & Rank \\ [0.5ex] 
		\hline 
		1 & T1 & 0.012 & 7 \\ 
		2 & T2 & 0.02 & 6 \\
		3 & T3 & 0.05 & 4 \\
		4 & T4 & 0.19 & 2 \\
		5 & T5 & 0.57 & 1 \\
		6 & T6 & 0.1 & 3 \\ 
		7 & T7 & 0.04 & 5 \\ 
		8 & T8 & 0.01 & 8 \\
		9 & T9 & 0.008 & 9 \\[1ex] 
		\hline 
		  & Tx Beam Entropy & 0.59 &  \\[1ex] 
		\hline 
	\end{tabular}
	\label{Tx Beam Rankings} 
\end{table}

\begin{table}[ht]
	\caption{ Rx Beam PMF and rankings} 
	\centering 
	\begin{tabular}{c c c c} 
		\hline\hline 
		Sl.\# & Beam\# & Probability & Rank \\ [0.5ex] 
		\hline 
		1 & R1 & 0.2 & 2 \\ 
		2 & R2 & 0.75 & 1 \\
		3 & R3 & 0.05 & 3 \\ 		[1ex] 
			\hline 
		& Rx Beam Entropy & 0.3 &  \\[1ex] 
		\hline 
	\end{tabular}
	\label{Rx Beam Rankings} 
\end{table}

\begin{figure}[!t]
	\hspace{-3pt}
	\includegraphics[width=0.99\columnwidth]{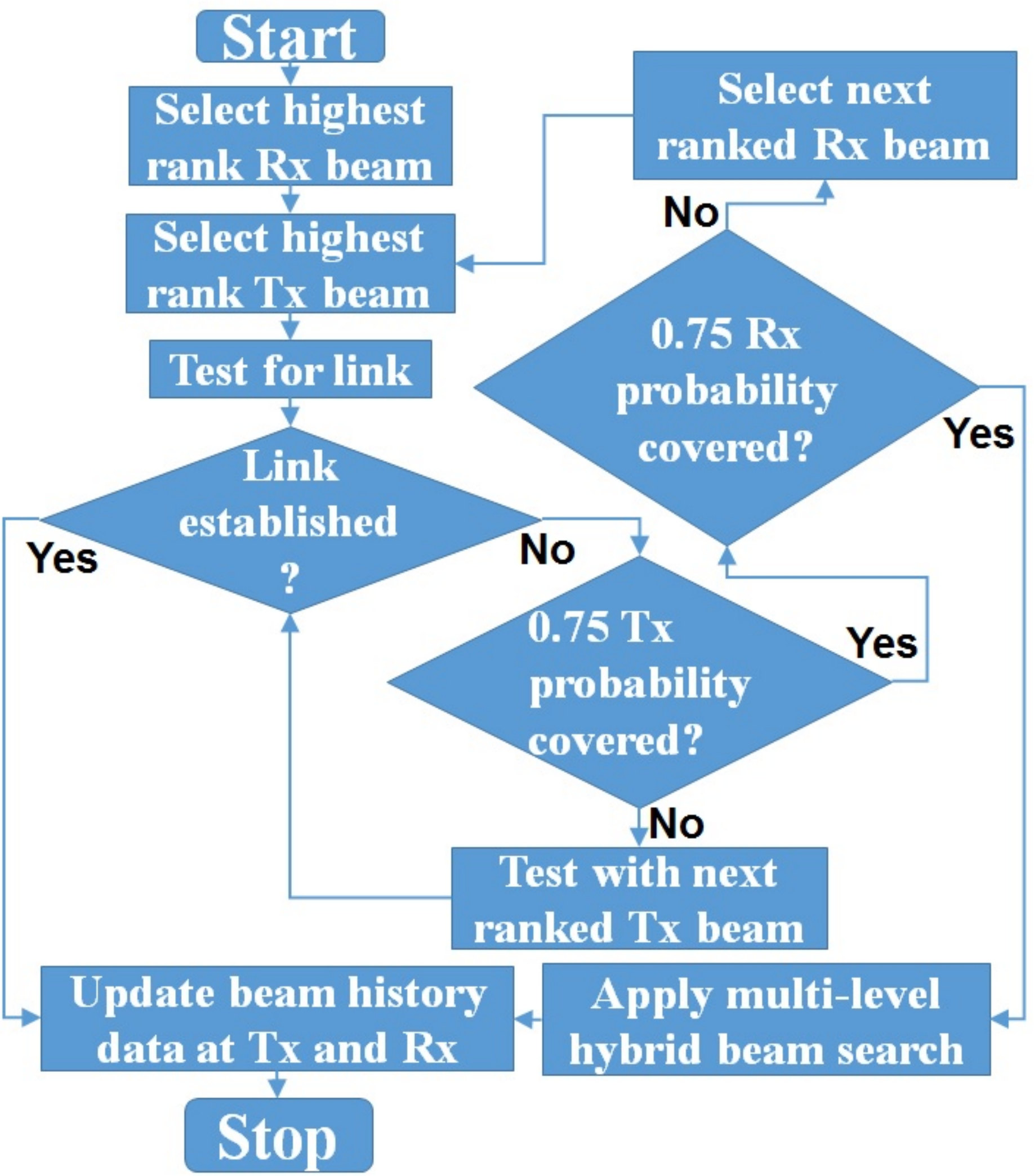}
	\caption{Pseudo-code flow diagram for MarS beam search algorithm}
	\label{my_alg}
\end{figure}

We propose a novel RF beam training algorithm for learning sparse MIMO channels in beamspace. The algorithm is depicted also as a flowchart in Figure \ref{my_alg} for easy readability. We assume the Tx and Rx beam ranking tables to be available at Tx and/or Rx prior to execution of the algorithm. If the Rx beam entropy is higher than the Tx beam entropy, then the Tx and the Rx shall be swapped in the algorithm. The lower entropy beam, e.g., Rx or Tx, is selected first for higher success probabilities in smaller number of tests. If the Tx and Rx ranking tables are at separate locations, then a minimal signaling may be required, which will be avoided if both the ranking tables are at each location. It also helps to reduce signaling requirements if Tx and Rx are synchronized and use uniform duration of time for testing a given Tx-Rx beam pair. In the second and third decision boxes of the flowchart in Figure \ref{my_alg}, a threshold value of 0.75 has been chosen for Tx/Rx cumulative probabilities for our example such that the beam training has statistically minimum overheads, i.e., on an average, a successful beam selection takes minimum number of beam tests. The threshold cumulative probability will be different for different PMFs. For less than 7 beams, having tested some of the beams and with ML being unfavorable, the remaining beams can be tested in the order of their probability-based ranks. For larger number of beams, a value for the threshold cumulative probability is chosen such that the remaining beams have individual probabilities less than 0.1 because ML will converge faster than many fine beams with extremely low success  probabilities. Every beam combination is tested only once. The failed beam combinations should be kept in memory / cache, at least until the successful beam training, so that they need not be tested again because it is already known that those Tx-Rx beam combinations do not correspond to a MIMO channel MPC. ML hybrid beam search is presented in the next section. The MarS RF beam training algorithm is presented below as Algorithm 1:

\RestyleAlgo{boxruled}
	\begin{algorithm}[ht]
		\caption{MarS beam training algorithm\label{algmars}}
		Select the highest rank Rx beam\\
		Select the highest rank Tx beam\\
		Test comm. link with selected Rx \& Tx beams\\
		~~~If link is established\\
		~~~~Go to Step 20\\
		~~~~~If cumulative Tx probability has not\\~~~~~ reached $0.75$\\
		~~~~~~~Select next ranked Tx beam\\
		~~~~~~~Go back to Step 3\\
	   ~~~~~Else if cumulative Tx probability has\\~~~~~ reached $0.75$\\
	   ~~~~~~~If cumulative Rx probability \\~~~~~~~has reached $0.75$\\
	   ~~~~~~~~~Apply ML hybrid beam search\\
	   ~~~~~~~~~Go to Step 20\\
	   ~~~~~~~Else if cumulative Rx probability\\~~~~~~~has not reached $0.75$\\
	   ~~~~~~~~~Select next ranked Rx beam\\
	   ~~~~~~~~~Go back to Step 2\\
	   Update beam history data for Rx and Tx\\
	   Stop  
	\end{algorithm}

\begin{table}[ht]
	\caption{MarS Success Probabilities} 
	\centering 
	\begin{tabular}{c c c c} 
		\hline\hline 
		Sl.\# & \# of tests & Success Probability \\ [0.5ex] 
		\hline 
		1 & 1 & 0.4275  \\ 
		2 & 2 & 0.1425  \\ 	[1ex] 
		\hline 
	\end{tabular}
	\label{aper} 
\end{table}	  

The respective probabilities of success for the first two tests are given in Table \ref{aper}. For example, success probability in one RSSI test is the joint probability of R2 and T5 and is equal to $0.75 \times 0.57= 0.4275$. Similarly, the probability of successful link establishment or Tx-Rx beam identification in the second test for R2 and T4 is $ 0.75 \times 0.19 = 0.1425$. Therefore, $57 \%$ of Tx-Rx beam pair identifications should be done in average $(0.4275 \times 1 + 0.1425 \times 2) \div (0.4275+0.1425) = 1.25$ beam tests. Also, we may say that $57 \%$ of the Tx-Rx beam pairs will be identified within respective first two tests. The success probabilities within a few initial tests increase with decrease in beam entropy, i.e., with decrease in channel randomness. Also, since high gain and narrow beams are used with all antenna elements being active, beam training can be performed at larger ranges than ML beam search in which initial levels use less number of active antenna elements for broader, but lower gain, beams.
 
 Let one beam training operation be the identification of one successful Tx-Rx beam pair for a communication link and  $1 \times 10^6$ beam training operations be performed during the life time of a communication system. For our example, exhaustive beam search performs $9 \times 3 = 27$ beam tests for one beam training operation and so $27 \times 10^6$ tests for the lifetime. For two levels at Tx only, pure ML beam search conducts $3 \times 3 + 3 \times 1 = 12$ beam tests for one beam training operation and $12 \times 10^6$ tests for the lifetime. MarS with ML hybrid beam search will execute a smaller average number of beam tests per beam training operation than pure MarS, i.e., without ML and only with ranked beam search. However, for simplicity let us consider pure MarS for a performance comparison. Let us define a mathematical operator `op' on ranked Rx and Tx PMFs, i.e., with the probabilities arranged in the descending order, such that the resultant vector is obtained by multiplying the entries of their Kronecker product by their respective position numbers in the Kronecker product vector, counting positions from `1'. Then, the average number of beam tests per Tx-Rx beam pair identification is given by summing the resultant vector elements and for our example we get $4.7$ as the answer. Let $\boldsymbol{r}$ and $\boldsymbol{t}$ denote ranked Rx and Tx PMFs, respectively, then the `op' mathematical operation consists of equation (\ref{kro}) and then equation (\ref{op}),
 
  \begin{equation}
 \label{kro}
 \centering
 \boldsymbol{k}= \boldsymbol{r} \otimes \boldsymbol{t},
 \end{equation}
 
\begin{equation}
\label{op}
\centering
\boldsymbol{x}= \boldsymbol{k} \odot [1:\text{length}(\boldsymbol{k})],
\end{equation}
where $\otimes$, length$(\boldsymbol{k})$, $\odot$,  and $\boldsymbol{q}=[1:n]$ represent Kronecker product, the number of elements in vector $\boldsymbol{k}$, Hadamard product, and a vector $\boldsymbol{q}$ given by first $n$ natural numbers from $1$ to $n$, respectively. The average number of beam tests per Tx-Rx beam pair identification is given by $m$ in equation (\ref{ave}),

\begin{equation}
\label{ave}
\centering
m= \text{sum}(\boldsymbol{x}),
\end{equation}
where sum({$\boldsymbol{x}$) denotes the sum of all elements in the vector $\boldsymbol{x}$. The number $m$ gives the average number of beam tests for identification of the first MPC Tx-Rx beam pair combination. For subsequent MPC Tx-Rx beam pair identifications, in cases of multi-MPC sparse MIMO channels, the average number of beam tests per MPC identification, i.e., per usable Tx-Rx beam pair identification, will be even smaller because every Tx-Rx beam pair is tested only once and the failed Tx-Rx beam pair combinations are saved in memory / cache until the completion of beam training for all the MPCs of the MIMO channel to be used for communications.

If Rx and Tx are swapped in the algorithm based on their entropies, then they will be swapped for the `op' operation also. As summarised in Table \ref{MarSper}, we note that for the first MPC beam training operation pure MarS, in the specific example, saves beam training overheads by $82.6 \%$ and $60.8 \%$ against exhaustive beam search for larger ranges and ML beam search for shorter ranges, respectively. 
 
 \begin{table}[ht]
 	\caption{MarS Performance comparison} 
 	\centering 
 	\begin{tabular}{c c c c} 
 		\hline\hline 
 		Sl.\# & Method & \# of Lifetime tests \\ [0.5ex] 
 		\hline 
 		1 & Exh. & $27 \times 10^6$  \\ 
 		2 & ML & $12 \times 10^6$  \\ 
 		3 & MarS & $4.7 \times 10^6$  \\ 	[1ex] 
 		\hline 
 	\end{tabular}
 	\label{MarSper} 
 \end{table}	
 
\section{Hybrid Beam Training Algorithm} 
\label{hybrid}
 For high beam entropy cases, i.e., for less uneven PMFs, a hybrid beam search method involving both MarS and ML techniques can be employed for large number of beams. In the hybrid scheme, we use multiple levels of statistically-ranked beam searches in which the later-level narrower beams are combined as earlier-level broader beams such that the beam entropies are as low as possible, especially for the first level.  
 
 \begin{table}[ht]
 	\caption{High entropy Tx Beam PMF and rankings} 
 	\centering 
 	\begin{tabular}{c c c c} 
 		\hline\hline 
 		Sl.\# & Beam\# & Probability & MarS Rank \\ [0.5ex] 
 		\hline 
 		1 & T1 & 0.13 & 4 \\ 
 		2 & T2 & 0.072 & 7 \\
 		3 & T3 & 0.05 & 8 \\
 		4 & T4 & 0.12 & 5 \\
 		5 & T5 & 0.25 & 1 \\
 		6 & T6 & 0.14 & 3 \\ 
 		7 & T7 & 0.08 & 6 \\ 
 		8 & T8 & 0.15 & 2 \\
 		9 & T9 & 0.008 & 9 \\[1ex] 
 		\hline 
 		& Tx Beam Entropy & 0.87 &  \\[1ex] 
 		\hline 
 	\end{tabular}
 	\label{n1} 
 \end{table}

Let us consider a Tx PMF as shown in Table \ref{n1}. The Tx beam entropy  of $0.87$ is higher in Table \ref{n1} than that of $0.59$ in Table \ref{Tx Beam Rankings}. For a given number of beams, the beam entropy gets highest when all the beams are equiprobable. For $9$ beams, maximum value of beam entropy is $\approx 0.95$ and an entropy of $0.87$ corresponds to $91.2 \%$ of the maximum possible beam entropy value, i.e., to a relative beam entropy of $0.912$. Having a range from 0 to 1, the relative beam entropy value depends only on the angular randomness of the channel and can be used as its single parameter indication in beamspace. Also, the relative beam entropy value remains the same even if the base for the logarithm is changed in equation (\ref{entropy}). If MarS algorithm was applied for Table \ref{n1}, beams T1 and T8 would be tested before beam T4 while we note that most of the probability mass is still concentrated in T4, T5, and T6. We consider two levels of beam search at Tx only for simplicity of illustration. There are only three Rx beams. In cases of large Rx antenna arrays, multi-level statistically-ranked beam search, i.e., hybrid beam search can be employed at Rx as well. 
 

If Tx beams T1, T2, and T3 are combined as a broader beam T01; T4, T5, and T6 as T02; and T7, T8, and T9 as T03, then the Tx broad beam PMF for the first level RF beam training will be as in Table \ref{n2}. We observe that the beam entropy reduces from $0.87$ in Table \ref{n1} to $0.448$ in  Table \ref{n2}. A reduction in beam entropy enables faster beam training by using probability-based beam rankings. 
   \begin{table}[ht]
 	\caption{Level 1 Tx Broad Beam PMF and Rankings} 
 	\centering 
 	\begin{tabular}{c c c c} 
 		\hline\hline 
 		Sl.\# & Level 1 Broad Beam\# & Probability & Rank \\ [0.5ex]
 		\hline 
 		1 & T01 & 0.252 & E2 \\ 
 		2 & T02 & 0.51 & E1 \\		
 		3 & T03 & 0.238 & E3 \\[1ex] 
 		\hline 
 		 & Level 1 Broad Beam Entropy & 0.448 &  \\[1ex] 
 		\hline 
 	\end{tabular}
 	\label{n2} 
 \end{table}	

Level 2 Tx beam PMFs and rankings will be as in Tables \ref{n3}, \ref{n4}, and \ref{n5} for level 1 beams T01, T02, and T03, respectively. 
 \begin{table}[ht]
	\caption{Level 2 Tx Beam Rankings for T01} 
	\centering 
	\begin{tabular}{c c c c} 
		\hline\hline 
		Sl.\# & Beam\# & Conditional Probability & Rank \\ [0.5ex]
		\hline 
		1 & T1 & 0.516 & 1Z1 \\ 
		2 & T2 & 0.283 & 1Z2 \\		
		3 & T3 & 0.201 & 1Z3 \\[1ex] 
		\hline 
	\end{tabular}
	\label{n3} 
\end{table}	
\begin{table}[ht]
	\caption{Level 2 Tx Beam Rankings for T02} 
	\centering 
	\begin{tabular}{c c c c} 
		\hline\hline 
		Sl.\# & Beam\# & Conditional Probability & Rank \\ [0.5ex]
		\hline 
		1 & T4 & 0.235 & 2Z3 \\ 
		2 & T5 & 0.49 & 2Z1 \\		
		3 & T6 & 0.275 & 2Z2 \\[1ex] 
		\hline 
	\end{tabular}
	\label{n4} 
\end{table}	
\begin{table}[ht]
	\caption{Level 2 Tx Beam Rankings for T03} 
	\centering 
	\begin{tabular}{c c c c} 
		\hline\hline 
		Sl.\# & Beam\# & Conditional Probability & Rank \\ [0.5ex]
		\hline 
		1 & T7 & 0.34 & 3Z2 \\ 
		2 & T8 & 0.63 & 3Z1 \\		
		3 & T9 & 0.03 & 3Z3 \\[1ex] 
		\hline 
	\end{tabular}
	\label{n5} 
\end{table}	

We propose the hybrid beam search algorithm as Algorithm 2. MarS would have identified a MPC corresponding to T4-R2 in 5 tests while hybrid beam training takes 4 tests. Also, $38.3 \%$ of the time, beam training will be done in only 4 tests despite a very high Tx beam entropy because the probability of T02-R2 is $0.38$.
	\begin{algorithm}[ht]
	\caption{Hybrid beam training algorithm\label{hybridalg}}
	Start with the first level beam search\\
	Select the highest rank Rx beam\\
	Select the highest rank Tx beam\\
	Test comm. link with selected Rx \& Tx beams\\
	~~~If link is established\\
	~~~~If final level is done\\
	~~~~~Update beam history data for/at Rx \& Tx\\
	~~~~~Stop  \\	
	~~~~Else if final level is  not done\\
	~~~~~Prepare for next level\\
	~~~~~Go to Step 2  \\
	~~~Else if link is not established\\
	~~~~If cumulative Tx probability has not\\~~~~reached $1$\\
	~~~~~Select next ranked Tx beam\\
	~~~~~Go back to Step 4\\
	~~~~Else\\
	~~~~~Select next ranked Rx beam\\
	~~~~~Go to Step 3	
\end{algorithm}

Tx/Rx beam PMFs will vary depending upon the VR programme, user characteristics, VR kiosk characteristics, etc. Therefore, there is no single family of PMFs which can be applied in general to Tx/Rx beam PMFs. In the rare case of relative beam entropy of `$1$' and a high beam entropy due to large number of beams, pure ML should be used. For smaller relative beam entropies, it will be beneficial to test the beams in the order of their probability-based ranks. If the link budget permits due to reduced beam gains for initial beam search levels, ML with probability-based ranking, i.e., hybrid beam search algorithm can be employed such that different levels of beam search have minimum possible beam entropies. Hybrid beam search algorithm enables the shortest possible training time, minimum overheads, and minimum communication losses due to receiver noise by reducing the average number of required beam tests for a successful Tx-Rx beam selection. However, the hybrid approach has the disadvantage that the probability of success will be null for the first $L-1$ tests where $L$ is the number of levels in the hybrid beam training algorithm. For larger communication distances with strict link budget conditions, pure MarS is useful. 

\section{Monte Carlo Simulations}
\label{matlab}

Monte Carlo simulations (MCS) were performed in Matlab for the example given at the end of Section \ref{mars} for MarS. Pure MarS was used to learn $1 \times 10^7$ random channel realizations created from the Tx beam and Rx beam PMFs. In Figure \ref{hist}, we see that $42.7 \%$ and $14.27 \%$ of Tx-Rx beam pairs were identified in only one and two beam tests, respectively, which is a very good agreement with Table \ref{aper}. We also note from MCS that overall only 4.7 beam tests are performed, on an average, per Tx-Rx beam pair identification, which is a saving of about $82.6 \%$ and $60.8 \%$ as compared to exhaustive and ML beam searches, respectively. The MCS results validate the probability model based values of Tables \ref{aper} and \ref{MarSper} for MarS.

\begin{figure}[!t]
	\hspace{-15pt}
	\includegraphics[width=7in]{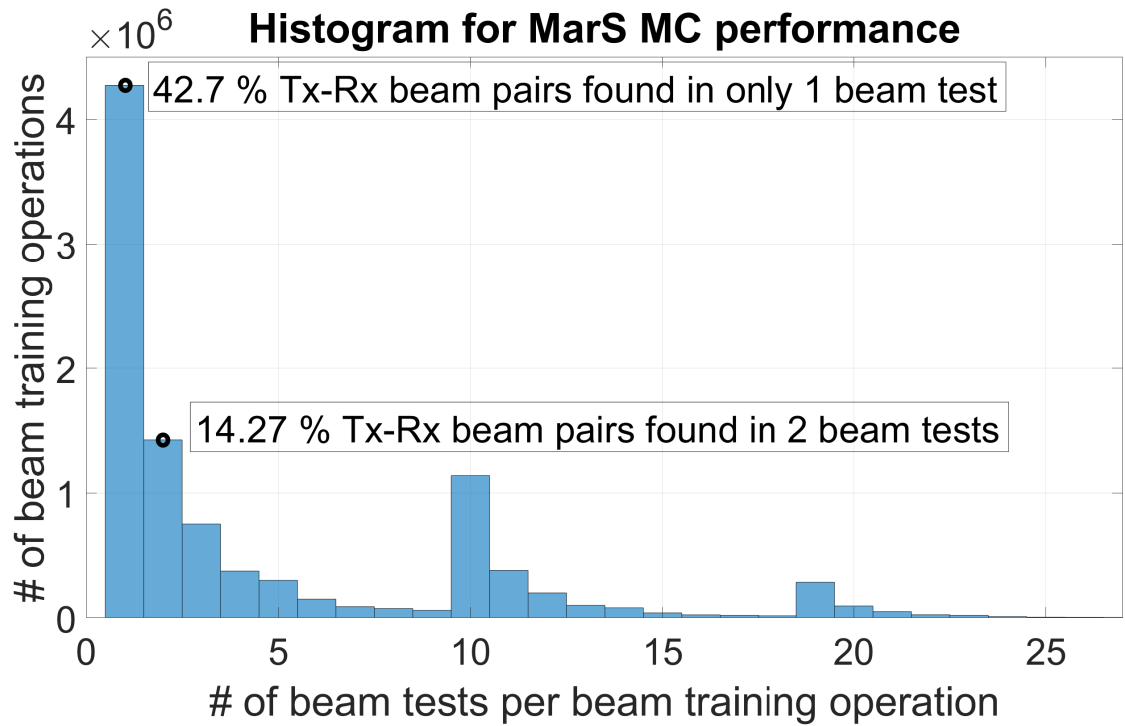}
	\caption{Histogram for MarS MC performance}
	\label{hist}
\end{figure} 

\section{Conclusions}
\label{conc}
Novel RF beam training algorithms have been proposed to learn sparse MIMO channels, especially for mm-wave and THz communication systems. In the MarS beam training algorithm, the beams are ranked based on their probabilities exploiting the memory-assisted statistical information inherent in the application specific scenario. The beam ranks can be updated depending upon the progressively available data. The beams are tested in the order of their ranks. The success probability is very high in the first few tests and increases with decrease in relative beam entropy. MarS reduces RF beam training time. 

For high entropy channels, a hybrid approach, using both MarS and ML beam search techniques, offers savings in beam training overheads and robustness against receiver noise.

In cases where no beam statistics is available, pure ML beam search can be employed initially. The beam  history data can be saved progressively with time to obtain the required statistics for either MarS or hybrid beam training. 

Although the usefulness of the proposed algorithms is convincing based on the probability model, MC simulation results for MarS have also been presented as a practical validation. 

It is planned to obtain typical beam entropy values for mm-wave channel models based on 3GPP report TR 38.901. In cases of correlation between Tx and Rx beams, the correlation can be used advantageously to further reduce training time and overheads. Such an investigation is an interesting future work.



 \section*{Acknowledgment}
 This work has received funding from the European Union's Horizon 2020 research and innovation programme under grant agreement No. 761329 (WORTECS).\\

The first author would like to thank Klaus Tittelbach-Helmrich and colleagues at IHP for reviewing the manuscript. 



%


\end{document}